\documentclass[preprint,letterpaper,sort&compress]{elsarticle}

\usepackage{natbib}
\usepackage{cancel}
\usepackage{pifont} 
\usepackage{amsthm}         
\usepackage{amsmath} 	   
\usepackage{amssymb}       
\usepackage{amsfonts}
\usepackage{graphicx} 	    
\usepackage{verbatim}
\usepackage{color}
\usepackage{colortbl}
\usepackage{float}
\usepackage{multirow}

\definecolor{babyblue}{rgb}{0.54, 0.81, 0.94}
\definecolor{corn}{rgb}{0.98, 0.93, 0.36}
\newcommand{\ww}[0]{$(w_0,w_a)$ }

\begin{document}

\begin{frontmatter}

\title{Assessing observational constraints on dark energy}

\author[a]{David Shlivko}
\ead{dshlivko@princeton.edu}

\author[a,b]{Paul J. Steinhardt\corref{cor1}}
\cortext[cor1]{Corresponding author}
\ead{steinh@princeton.edu}

\address[a]{Department of Physics, Princeton University, Princeton, NJ 08544, USA}
\address[b]{Jefferson Physical Laboratory, Harvard University, Cambridge MA 02138, USA}

\date{\today}

\begin{abstract}
Observational constraints on time-varying dark energy ({\it e.g.}, quintessence) are commonly presented on a $w_0$-$w_a$ plot that assumes the equation of state of dark energy strictly satisfies $w(z)= w_0+ w_a z/(1+z)$ as a function of the redshift $z$.  Recent observations favor a sector of the $w_0$-$w_a$ plane in which $w_0 > -1$ and $w_0+w_a< -1$, suggesting that the equation of state underwent a transition from violating the null energy condition (NEC) at large $z$ to obeying it at small $z$. In this paper, we demonstrate that this impression is misleading by showing that simple quintessence models satisfying the NEC for all $z$ predict an observational preference for the same sector. 
We also find that quintessence models that best fit observational data can predict a value for the dark energy equation of state at present that is significantly different from the best-fit value of $w_0$ obtained assuming the parameterization above. 
In addition, the analysis reveals an approximate degeneracy of the $w_0$-$w_a$ parameterization that explains the eccentricity and orientation of the likelihood contours presented in recent observational studies. 
 \end{abstract}


\end{frontmatter}

\section{Introduction}
Models of time-varying dark energy ({\it e.g.}, quintessence) \cite{Ratra:1987rm,Peebles:1987ek,Wetterich:1987fm,Coble:1996te,Turner:1997npq,Caldwell:1997ii} can  be  constrained  by combining observations at different redshifts, such as the cosmic microwave background (CMB), baryon acoustic oscillations (BAO), and Type IA supernovae (SNe Ia) \cite{Planck:2018vyg,DESI:2024mwx,Brout:2022vxf,Rubin:2023ovl,DES:2024tys}. A common convention is to display the observational constraints on a $w_0$-$w_a$ plot assuming the variation of the dark energy equation of state  $w$ is well-described by a two-parameter function of the redshift $z$ that takes the Chevallier-Polarski-Linder form \cite{Linder:2002et,chevallier_accelerating_2001}
\begin{equation}\label{wz}
w(z)= w_0 + w_a z /(1+z).	
\end{equation}
Recent observations \cite{DESI:2024mwx,Brout:2022vxf,Rubin:2023ovl,DES:2024tys} appear to favor a sector of the $w_0$-$w_a$ plane with $w_0 > -1$ and $w_0+w_a< -1$, in which case Eq. (\ref{wz}) predicts that the null energy condition (NEC) and weak energy condition (WEC) are violated at large redshift ({\it i.e.}, $w(z) < -1$) but satisfied at small redshift ($w(z) \geq -1$).

One must be careful when portraying observational results in this way, because cosmological observables depend directly on the Hubble parameter $H(z)$ and its integrals but only indirectly on the dark energy equation of state $w(z)$. In particular, the functional form of $w(z)$ can differ significantly from Eq. (\ref{wz}) in different models of quintessence, while still predicting very similar cosmological observables \cite{Linder:2002et,dePutter:2008wt,garcia-garcia_theoretical_2020}.
In order to compare the predictions of  different quintessence models to one another and to observations using a $w_0$-$w_a$ plot, one can assign to each model the ordered pair $(w_0, \, w_a)$ that predicts an $H(z)$ most similar to that of the quintessence model.  This is similar to the approach suggested by de Putter and Linder \cite{dePutter:2008wt}, who showed that even just by assigning $w_a$ in this way, it is possible to match $H(z)$ to $\sim 0.1\%$ or better for some models of quintessence.   

In this paper, we use this mapping technique to predict where observational preferences should fall on the $w_0$-$w_a$ plane for a range of quintessence models driven by a scalar field $\varphi$ with canonical kinetic energy density and potential $V(\varphi)$.   The models we consider, sometimes referred to as ``thawing dark energy''  \cite{Scherrer:2007pu},  have the property that  $w(z)\rightarrow -1$ at early times (large $z$) because Hubble friction during the matter- and radiation-dominated eras is large enough to freeze the scalar field.  At late times (small $z$),  the Hubble friction becomes negligible,  the scalar field accelerates down the potential, and $w(z)$ increases as $z$ decreases. We show that,  when matching $H(z)$'s, these models are mapped onto the same sector of the $w_0$-$w_a$ plot as observations currently prefer, even though they do not violate the NEC.  As a result, observations favoring a region with  $w_0+ w_a < -1$ do not imply that dark energy must be NEC-violating at any redshift. 

We detail our mapping protocol in Sec. \ref{methods} and discuss the uncertainties in this procedure (due partially to an approximate degeneracy in the $w_0$-$w_a$ plane) in Sec. \ref{uncertainty}. The quintessence models we will test are presented in Sec. \ref{models}, and the results of their mapping onto the $w_0$-$w_a$ plane are given in Sec. \ref{results}. Finally, in Sec. \ref{discussion}, we summarize our findings and discuss the implications for interpreting observational likelihood contours on a $w_0$-$w_a$ plane. In particular, we point out how the observations bear on issues such as NEC violation, consistency with supergravity and string theory, and on whether accelerated expansion continues forever or terminates and transitions to contraction.

\section{Methods}\label{methods}
As noted above, our procedure relies on determining which combination \ww is the ``best fit'' to a given quintessence model of dark energy. Rather than choosing the fit that best mimics the quintessence field's equation of state $w_Q(z)$, we choose the fit that best matches the evolution of the Hubble parameter $H_Q(z)$ in the quintessence model. This allows the fit to most nearly reproduce key late-time cosmological observables, such as the Hubble distance
\begin{equation}
	D_H(z) \equiv 1/H(z),
\end{equation}
the angular diameter distance
\begin{equation}
	D_M(z) \equiv \int_0^z H(\tilde{z})^{-1}d\tilde{z},
\end{equation}
and the luminosity distance
\begin{equation}
	D_L(z) = (1+z)^2D_M(z).
\end{equation}
Notably, when optimizing to match $H(z)$, the integrated quantities $D_M(z)$ and $D_L(z)$ are found to match with similar accuracy, as we will show for a sample model in Sec. \ref{results}.

Quantitatively, we define the ``best fit'' combination $(w_0,w_a)$ as one that minimizes the error
\begin{equation}\label{error}
	E \equiv \max_{\text{z $ < $ 4}}\left|\frac{H_\text{fit}(z)-H_Q(z)}{H_Q(z)}\right|.
\end{equation}
Here, $H_Q(z)$ is the evolution of the Hubble parameter for a given quintessence model, while $H_\text{fit}(z)$ is the evolution for an artificial \ww model. The maximization in Eq. (\ref{error}) is performed over the finite interval $z < 4$ to roughly match the redshifts probed by BAO and SNe Ia measurements. These are also the redshifts during which the time-variation of dark energy is most relevant; at higher redshifts, the universe is strongly matter-dominated, and $H(z)$ can be computed directly from the present-day matter density $\rho_{m,0}$, independently of the nature or behavior of dark energy. Note that there is some flexibility to the above definition of error; for example, the approach in Ref. \cite{dePutter:2008wt} matches $D_M(z)$ over all $z > 0$ rather than $H(z)$ over $z < 4$. In Sec. \ref{uncertainty}, we will discuss how this flexibility translates to an uncertainty in our best-fit results. 

For some models of quintessence, $H_Q(z)$ can be easily calculated from analytically parameterized expressions for $w_Q(z)$ (see, \emph{e.g.}, Refs. \cite{Scherrer:2007pu,Dutta:2008qn}). However, these expressions are approximate and only valid in the regime where $1+w_Q(z) \ll 1$, which limits the range of models that can be studied. In this work, in order to find $H_Q(z)$ for a quintessence model driven by a canonical scalar field $\varphi$ with potential $V(\varphi)$,
we switch variables from $z$ to $N \equiv -\ln(1+z)$ and numerically solve the full equations of motion
\begin{align}
	\varphi''(N) &= \frac{3}{2}\varphi'(N)\left(w_Q(N)\Omega_Q(N)-1\right) - \frac{V_{,\varphi}}{H_Q(N)^{2}}, \\
	\Omega_Q'(N) &= 3w_Q(N)\Omega_Q(N)(\Omega_Q(N)-1), \\
	w_Q(N) &= \frac{H_Q(N)^2\varphi'(N)^2}{V(\varphi) + \frac{1}{2}H_Q(N)^2\varphi'(N)^2}-1, \\
	H_Q(N) &= \sqrt{\frac{V(\varphi)}{3\Omega_Q(N)-\frac{1}{2}\varphi'(N)^2}}.
\end{align}
Here, $V_{,\varphi} \equiv dV(\varphi)/d\varphi$, $w_Q(N)$ is the equation of state of the scalar field, $\Omega_Q(N) = 1-\Omega_m(N)$ is the fraction of total energy density attributable to the scalar field, and primes denote derivatives with respect to $N$. Note that these equations assume a spatially flat universe. We set our initial conditions deep in the matter-dominated past, with $\Omega_{Q} = 10^{-6}$. The initial field value will depend on the particular model, but the field velocity $\varphi'(N)$ can be initially set to zero without loss of generality, as it will evolve toward an attractor trajectory while $\Omega_Q \ll 1$. We end the simulation upon reaching a pre-selected fiducial value of $\Omega_Q(0)$, which we take to be an input to the mapping procedure. Note that throughout this work, we will measure $H_Q(N)$ in units of $H_Q(0)$, and in the examples we provide, we will assume that $\Omega_Q(0) = 0.7$.

Next, we can generate a candidate fit $H_\text{fit}(N)$ by solving the system
\begin{align}
	\Omega_\text{fit}'(N) &= 3w_\text{fit}(N)\Omega_\text{fit}(N)(\Omega_\text{fit}(N)-1), \\
	H_\text{fit}'(N) &= -\frac{3}{2}H(N)(1+w_\text{fit}(N)\Omega_\text{fit}(N)), \\
	w_\text{fit}(N) &\equiv w_0 + w_a (1-e^N).
\end{align}
Here, we do \emph{not} assume that the fitted fractional dark energy density $\Omega_\text{fit}(0)$ equals $\Omega_Q(0)$, nor do we assume that $H_\text{fit}(0)$ equals $H_Q(0)$. Instead, we treat these two initial values as free parameters, in addition to the combination \ww specifying the equation of state $w_\text{fit}(N)$. We do check, however, that the best-fit values for $\Omega_m(0) \equiv 1-\Omega_\text{fit}(0)$, $H_\text{fit}(0)$, and the combination $\Omega_m(0)H_\text{fit}(0)^2$ are within $\sim 1\%$ of their values in the quintessence model. This ensures that the fit is compatible with CMB constraints in addition to measurements of the low-redshift universe.

The final step in our procedure is to scan over the four free parameters and identify the best-fit combination of $\{w_0, w_a, \Omega_\text{fit}(0), H_\text{fit}(0)\}$ with the smallest error as defined in Eq. (\ref{error}). We note that fitting all four parameters, rather than just $w_a$ as in the approach of de Putter and Linder, is necessary in order to properly identify the best fits for a wider range of quintessence models than those considered in Ref. \cite{dePutter:2008wt}. After projecting out the best-fit values of $H_\text{fit}(0)$ and $\Omega_\text{fit}(0)$, we are left with a single point on the $w_0$-$w_a$ plane that best represents the quintessence model we started with. More generally, this protocol can be used to map a one-parameter family of quintessence models onto a best-fit curve on the $w_0$-$w_a$ plane, or a multi-parameter family of models onto a best-fit region. 
 
 \section{Degeneracy and uncertainty}\label{uncertainty}
 
  \begin{figure}[t]
 \begin{center}
\includegraphics[width=0.75\textwidth]{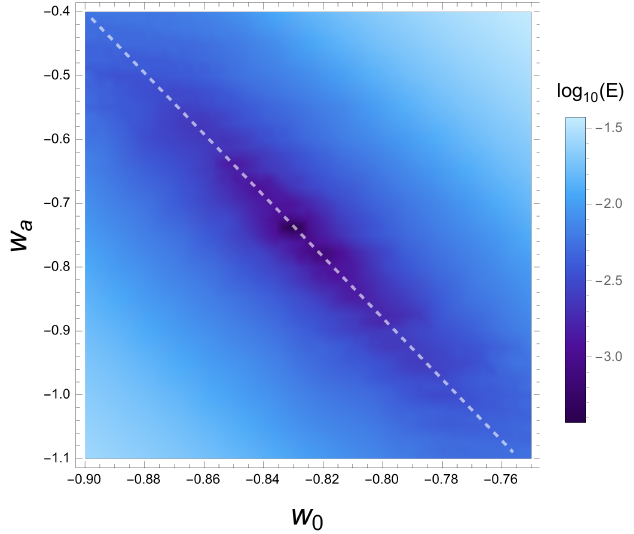}
\caption{\label{wwErr} A mild degeneracy among \ww models is illustrated by computing the error $E$ of each \ww fit when matching the $H(z)$ of a fiducial \ww model with $w_0 = -0.827, w_a = -0.75$. The orientation of the degeneracy is indicated by the dashed line with slope $\Delta w_a/\Delta w_0 = -4.8$.}
\end{center}
\end{figure}  

In our fitting process, it is instructive to identify a region of ``acceptable'' combinations of \ww in addition to the central or ``best-fit'' combination. There is an approximate degeneracy between $w_0$ and $w_a$ when matching cosmological observables, causing this acceptable region to be highly eccentric. In particular, for any best-fit combination \ww that matches a quintessence model's $H(z)$ with reasonable accuracy, there will be a large set of combinations satisfying $\Delta w_a/\Delta w_0 \approx -5$ that match the model with similar accuracy. The slope of this degeneracy can vary by $\sim 10\%$ depending on the model being fitted, but its qualitative appearance is clear and distinct. 

To demonstrate that this effect is fundamental to the \ww parameterization and not dependent on any particular choice of quintessence model, Fig. \ref{wwErr} illustrates the degeneracy when fitting \ww combinations to a fiducial model strictly obeying the parameterized equation of state in Eq. (\ref{wz}) with \ww = $(-0.827, -0.75)$; these are the central values of the DESI+CMB+PantheonPlus constraints \cite{DESI:2024mwx}. The darkest regions in the figure correspond to the \ww combinations that most accurately match the fiducial model's $H(z)$, and the dashed line with slope $\Delta w_a/\Delta w_0 = -4.8$ indicates the axis of degeneracy. One can see that many combinations aligned along this axis are accurate fits to the central model within $E \lesssim 0.3\%$. 

This degeneracy is not only relevant for our theoretical fits to quintessence models, but it also makes a prediction about the orientation of constraint contours in the $w_0$-$w_a$ plane produced by observational analyses. In particular, if the observed values of $H(z)$ over a broad range of redshifts are well fit by one combination of ($w_0, w_a$), then we expect that it will also be reasonably fit by other combinations along the axis of degeneracy. The resulting eccentricity and orientation of the constraint contours are indeed apparent in, \emph{e.g.}, Refs. \cite{DESI:2024mwx,Brout:2022vxf,Rubin:2023ovl,DES:2024tys}, with the slope most closely matching our prediction in the combined BAO+CMB+SNe Ia results from DESI \cite{DESI:2024mwx}.

 \begin{figure}[H]
 \begin{center}
\includegraphics[width=0.75\textwidth]{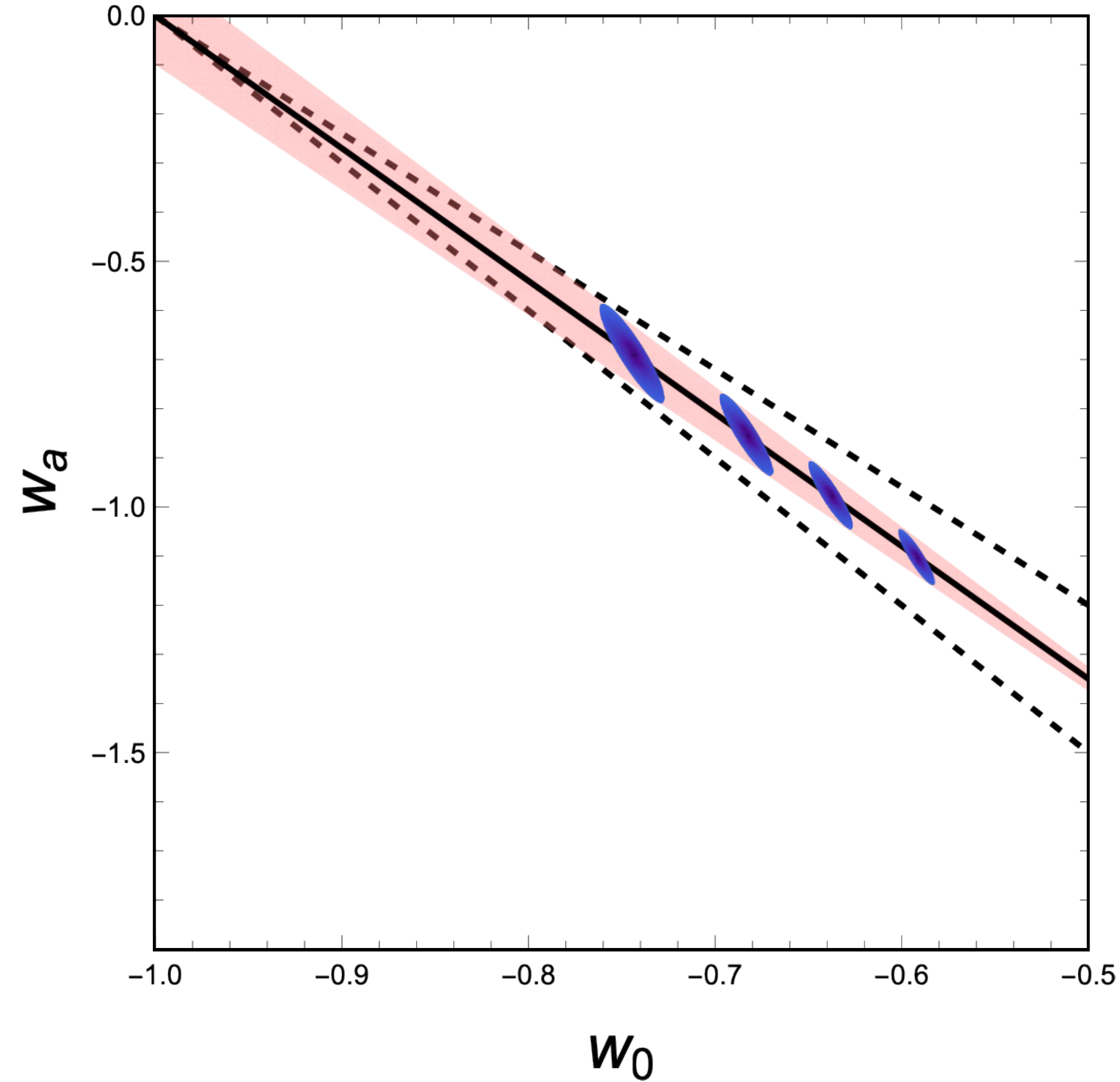}
\caption{\label{uncplot} Uncertainties associated with the fitting protocol. A line of best-fit \ww combinations for a hypothetical one-parameter quintessence model is shown in solid black. Each point on the line has an associated ``acceptable-fit'' region of variable size, depicted here qualitatively as blue elliptical contours satisfying $E \leq E_\text{max}$ for some $E_\text{max}$. The contours are generically largest near the $\Lambda$CDM limit, and they blend together to form a tapered acceptable-fit ribbon (shaded pink) around the best-fit line. A second type of uncertainty, representing the flexibility in the definition of error when determining the best-fit line itself, is illustrated qualitatively with two alternative best-fit lines (dashed) converging in the $\Lambda$CDM limit.}
\end{center}
\end{figure}

Unlike the orientation of the degeneracy, the size of the degenerate region in theoretical \ww fits is model-dependent. In the example shown in Fig. \ref{wwErr}, the best-fit error is $E = 0$ by construction, and there is a correspondingly large region of acceptable fits, depending on the precision $E_\text{max}$ with which $H(z)$ can be measured. In general, a larger error $E$ for the best-fit \ww combination corresponds to a smaller region of acceptable fit.

When plotting a best-fit curve on the \ww plane corresponding to a one-parameter family of quintessence models, the acceptable-fit regions of each point merge together to form an acceptable-fit ribbon, shown in Fig. \ref{uncplot} as the pink shaded region. The ribbon is widest at the $\Lambda$CDM limit ($w_0 = -1, w_a = 0$) of any quintessence model, provided such a limit exists, but the rate at which the ribbon tapers off is model-dependent. 

Finally, a second and entirely independent type of uncertainty reflects our confidence (or lack thereof) in the location of the best-fit curve itself on the $w_0$-$w_a$ plane. This second uncertainty stems from a fundamental ambiguity in the notion of ``best-fit,'' \emph{i.e.}, in the definition of error (Eq. \ref{error}). We have observed that choosing a different definition of error---say, a mean-square error instead of a maximum, or matching $D_M$'s instead of $H$'s---can affect the slope of best-fit lines by $\lesssim 10\%$, depending on the model. The lines pivot about the $\Lambda$CDM limit (assuming such a limit exists in the models being fitted), where all definitions of error agree that the cosmological observables are best fit by \ww$ = (-1, 0)$. This uncertainty in the slope is depicted by the dashed lines in Fig. \ref{uncplot}.

\section{Models}\label{models}
In this work, we illustrate the methods outlined in the previous section using three classes of thawing quintessence models as examples: exponential potentials, hilltops, and plateaus. Each of these models is driven by a scalar field $\varphi$ with canonical kinetic energy density rolling down a potential $V(\varphi)$. The scalar field is initially held constant during the radiation- and matter-dominated epochs by Hubble friction, such that $w \to -1$ at large redshift.  Then, as the dark energy density becomes comparable to the matter density, the Hubble friction decreases, the field accelerates down the potential, and in turn $w$ increases as $z$ decreases (or ``thaws'' away from $-1$). This behavior corresponds to a negative value of $w_a$ in the \ww parameterization, as appears to be preferred by recent observational constraints \cite{DESI:2024mwx,Brout:2022vxf,Rubin:2023ovl,DES:2024tys}. 

We parameterize the exponential potential as
\begin{equation}
	V_\text{exp}(\varphi) = V_0e^{\lambda\varphi},
\end{equation}
where $\varphi$ is in units of the reduced Planck mass $M_{pl} = \sqrt{\hbar c / (8\pi G)}$ and the initial field value is set to zero, such that $V_0 \sim \mathcal{O}(H_0^2M_{pl}^2)$. A theoretical motivation for studying this class of models is that scalar fields with exponential potentials are ubiquitous in supergravity, modified gravity, and superstring theories; see for example \cite{Wetterich:1994bg,Binetruy:1998rz,Bedroya:2019snp}. Additionally, the exponential potential probes two interesting limits of quintessence models: one where $V_{,\varphi}/V$ is constant and much smaller than $M_{pl}^{-1}$ (``slow-roll'' thawing quintessence \cite{Scherrer:2007pu,Chiba:2009sj}), and one where $V_{,\varphi}/V$ is constant but not small (\emph{i.e.}, when $\lambda \gtrsim M_{pl}^{-1}$). 

Hilltop potentials with large, negative $V_{,\varphi\varphi}$ provide a complementary probe into the regime where $V_{,\varphi}/V$  is small compared to $M_{pl}^{-1}$ but not constant.  Potentials of this type are good approximations to axion models (or pseudo-Nambu Goldstone bosons generally) provided the scalar field is initially frozen by Hubble friction near the top of its potential \cite{Dutta:2008qn}.  
We take a quadratic approximation and write the hilltop potential as 
\begin{equation}
	V_\text{hill}(\varphi) = V_0\left(1 - \frac{1}{2}k^2\varphi^2\right),
\end{equation}
where again $\varphi$ is in units of $M_{pl}$ and $V_0 \sim \mathcal{O}(H_0^2M_{pl}^2)$. In this work, we analyze a relatively flat hilltop with $k^2M_{pl}^2 = 1$ and a more concave hilltop with $k^2M_{pl}^2 = 100$. For each case, we will use a variety of initial field values to map out the set of possible best-fit \ww combinations to the given potential.

The plateau potential we consider is something of a hybrid between the previous two examples, pairing a region of smooth exponential decay with a sharp, cliff-like drop:
\begin{equation}\label{hybrid}
	V_\text{plateau} = V_0\left(e^{-\varphi/M} - \kappa e^{\varphi/m}\right).
\end{equation}
This is an example of a case where $V_{,\varphi}/V$ cannot be assumed to be either small or constant. We choose an exponential cliff (rather than a power-law) to match the model of Ref.~\cite{Andrei:2022rhi}, where it was shown that the potential (\ref{hybrid}) can lead to a transition from accelerated expansion to a regime of slow contraction when $V_\text{plateau}$ becomes negative, as can occur in a cyclic cosmology. For this model, we are free to set the initial field value to $\varphi = 0$ without loss of generality. As before, $\varphi$ is in units of $M_{pl}$, $V_0 \sim \mathcal{O}(H_0^2M_{pl}^2)$, and $\kappa \lesssim 1$ is a constant. This model has three free parameters, $\kappa$, $M$, and $m$, which leads to a best-fit region on the $w_0$-$w_a$ plane rather than a curve. Note that in any limit where the second term in the potential becomes negligible, we recover the exponential model with $\lambda = 1/M$.

\section{Results}\label{results}
\begin{figure}[t]
 \begin{center}
\includegraphics[width=0.75\textwidth]{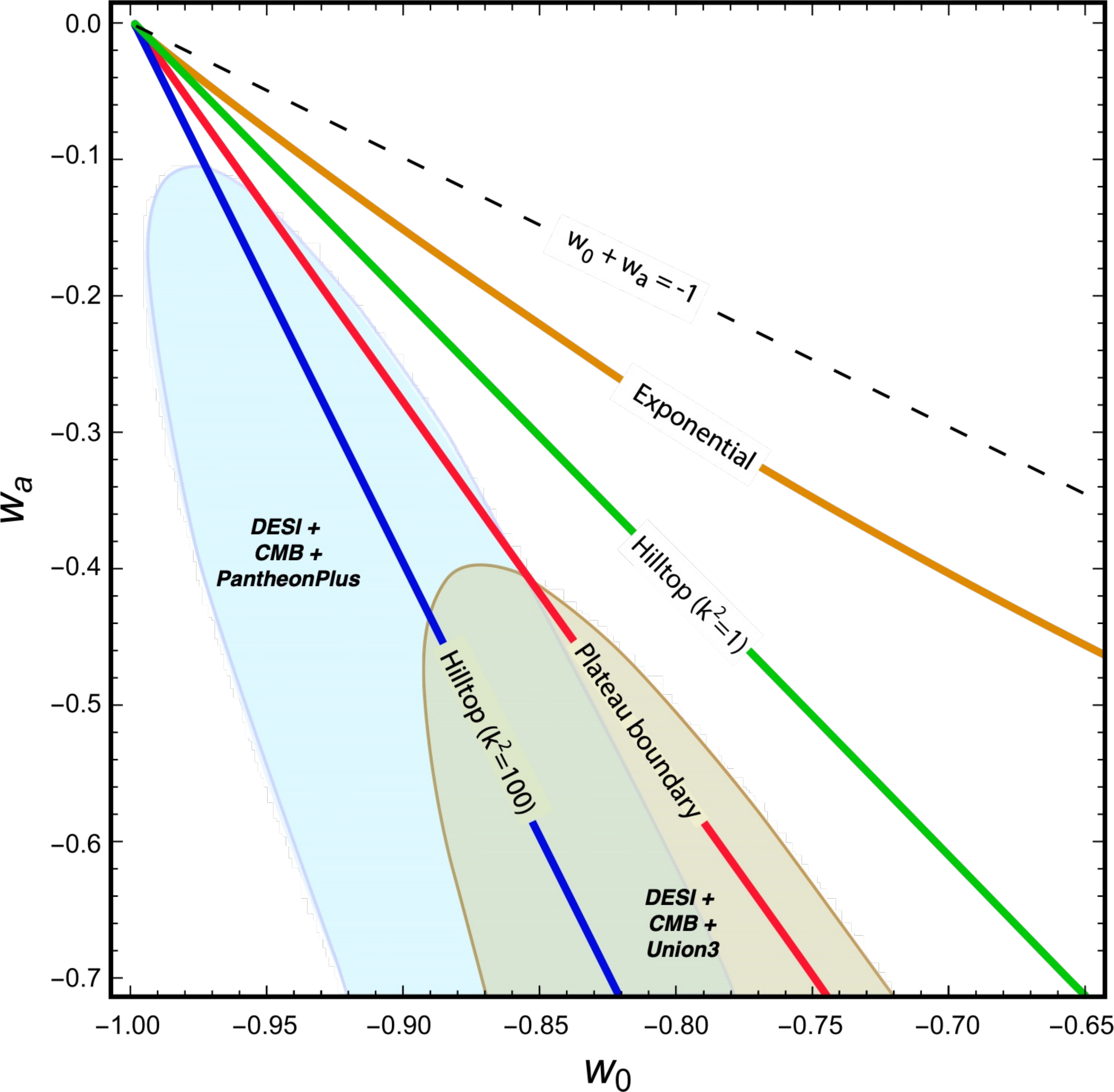}
\caption{\label{allLines} A $w_0$-$w_a$ plot showing the predictions for three types of canonical quintessence potentials (exponential, plateau, hilltop) that are NEC-satisfying. For plateau models, the predictions map onto the region between the exponential curve and the plateau boundary curve indicated. The plot also shows the best-fit contours from DESI BAO+CMB+(PantheonPlus or Union3) that prefer the sector  $w_0+ w_a < -1$  (below the dark dashed line), naively suggesting NEC violation at large $z$. }
\end{center}
\end{figure}

The best-fit curves and regions for the three classes of models introduced in Sec. \ref{models} are depicted in Fig. \ref{allLines}. As mentioned in Sec. \ref{methods}, this analysis assumes a fiducial value of $\Omega_Q(0) = 0.7$.

 The orange curve corresponds to the exponential model. The limit $\lambda \to 0$ maps onto the $\Lambda$CDM parameters $w_0 = -1$ and $w_a = 0$. Each fit along this curve has error $E \lesssim 0.1\%$, which is the greatest level of accuracy among the models we tested.

The region between the exponential curve (orange) and the plateau boundary (red) corresponds to the three-parameter family of plateau models.  The upper boundary (the exponential curve)  corresponds to the limit  $\kappa \ll 1$ in which the cliff is negligibly small.   The lower boundary corresponds to large $\kappa$ and a substantial cliff that causes a sudden increase in the equation of state as $z$ decreases. This sudden increases in $w(z)$ causes the models to be mapped onto more negative values of $w_a$. The errors near this boundary can be somewhat greater but still satisfy $E \lesssim 0.5\%$ for the cases we tested. We note that the lower boundary of this region is approximate, and other types of plateau models may not mimic the behavior of the specific potential examined in this work.

The two remaining curves correspond to hilltop models with $k^2M_{pl}^2 = 1$ (green) and $k^2M_{pl}^2 = 100$ (blue). The respective errors satisfy $E \lesssim 0.2\%$ and $E \lesssim 0.7\%$ within the region shown in the plot. The closer the initial field value is to the hilltop, the closer the best-fit \ww model is to $\Lambda$CDM, and the smaller is the best-fit error.

These results are overlaid with the observational constraints obtained by the DESI collaboration (blue and brown $2\sigma$ contours) when combining measurements of BAO, CMB, and SNe Ia \cite{DESI:2024mwx}. This allows the plot to be used for comparing quintessence models to each other and to the observations at the same time. For example, the likelihood contours in Fig. \ref{allLines} appear to favor quintessence potentials with sharp drops, like plateau models with the steepest cliffs or hilltop models with $k^2M_{pl}^2 \sim 100$. However, we make this point mainly for the purpose of illustrating the utility of $w_0$-$w_a$ plots in general, and we would not suggest drawing any strong conclusions based on currently available data.

Notably, \emph{all} the quintessence models we tested obey the NEC and have $w(z) > -1$ for all $z$, yet they are best fit by curves or regions on the $w_0$-$w_a$ plane that satisfy $w_0 + w_a < -1$. This boundary is marked by the dashed line in Fig. \ref{allLines}. We therefore conclude that a preference for this sector of the plane, which would naively imply NEC violation at large $z$ according to equation (\ref{wz}), is actually compatible with NEC-satisfying models of quintessence. This result is a manifestation of the fact that even large differences in $w(z)$ over a wide range of $z \gg 1$ (where $\Omega_Q(z)$ is negligibly small) have a negligible impact on cosmological observables such as $H(z)$.

 \begin{figure}[H]
 \begin{center}
\includegraphics[width=0.75\textwidth]{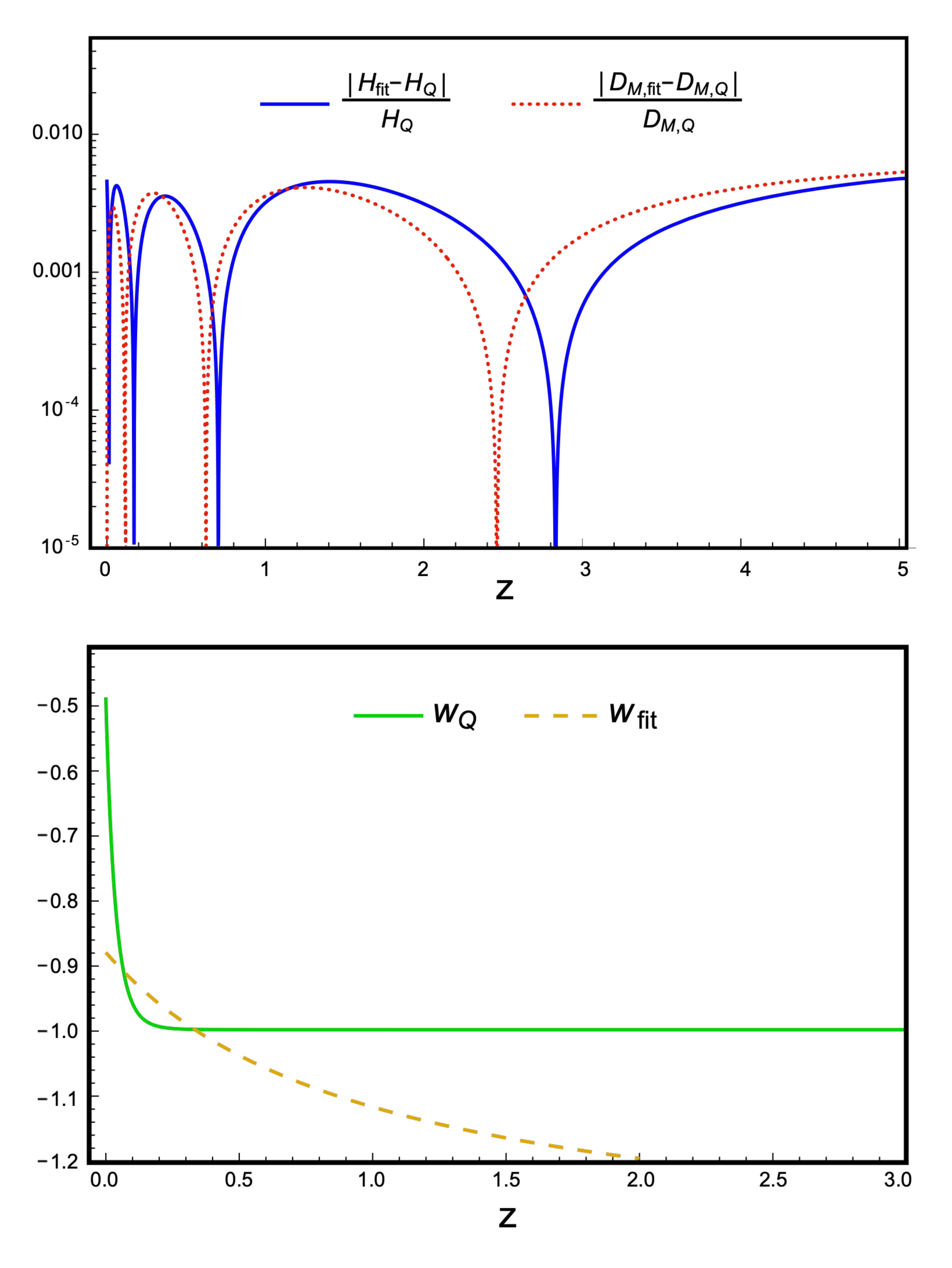}
\caption{\label{erp} [Top panel] Fractional difference in $H(z)$ and $D_M(z)$ between a hilltop quintessence model with $k^2 = 100$ and the best-fit \ww model. Note that $|\Delta D_L|/D_{L,Q} = |\Delta D_M|/D_{M,Q}$. For this model, as for all models considered in this work, the Hubble parameter $H(z)$ and the integrated quantities $D_M(z)$ and $D_L(z)$ are matched with sub-percent errors at all $z$. \newline [Bottom panel] The evolution of $w_Q(z)$ for the same hilltop model (solid curve), compared with the evolution of $w_\text{fit}(z)$ for the best-fit \ww combination (dashed curve) based on matching $H(z)$'s.}
\end{center}
\end{figure}  
 At smaller redshifts, $H(z)$ is more sensitive to differences in $w(z)$, but only through an integral. As a result, two models can have very similar $H(z)$'s even if their $w(z)$'s differ noticeably over a narrow range of redshifts near $z=0$. In turn, this means that the best-fit values of $w_0$ obtained when matching $H(z)$'s need not be equal to---or even close to---the actual value $w_Q(0)$ for the quintessence model being considered. This is specifically the case for the hilltop and plateau models, in which $w_Q(z)$ is increasing rapidly as $z \to 0$ (see, \emph{e.g.}, Refs. \cite{linden_test_2008,wolf_underdetermination_2023}). We illustrate an example of this behavior for a hilltop model with $k^2M_{pl}^2 = 100$ in Fig. \ref{erp}, juxtaposing the fits to $H(z)$ and $D_M(z)$, which have sub-percent level errors, with a comparison of $w(z)$'s between the model and the fit, which differ by $\mathcal{O}(100\%)$. In fact, the best-fit curves drawn in Fig. \ref{allLines} include cases where $w_Q(0) > -1/3$, in which case the universe today is no longer accelerating, while the best-fit \ww combination still has $w_0 \leq -0.65$.

As discussed in Sec. \ref{uncertainty}, there is some uncertainty in our best-fit results due to the flexibility in how we define the error $E$. This uncertainty is negligible for the exponential curve, but the plateau boundary and hilltop curves should be interpreted as having a $\sim 10\%$ error bar on the value of $w_a$ at any given $w_0$. Additionally, extending the plateau and hilltop curves beyond the region shown would produce increasingly large best-fit errors $E$, ultimately reaching a point where there exists no good fit to the quintessence model within the \ww parameter space. In cases like this---including more general models of dark energy or modified gravity that are not well fit by any \ww combination---the safer approach would be to perform a separate analysis specific to the model in question, rather than trying to map it onto, and then constrain, the $w_0$-$w_a$ parameter space.

Finally, we note that plots like Fig. \ref{allLines} carry no information about fine-tuning of either parameters or initial conditions for a given model. For example, in a hilltop model, only a small range of initial field values is simultaneously sufficiently close to the peak of the hilltop to produce a noticeable period of accelerated expansion and sufficiently far from the hilltop to be distinguishable from $\Lambda$CDM. This fine-tuning must be judged and weighed separately.

\section{Discussion}\label{discussion}

In this work, we modified and illustrated a mapping protocol that assigns a best-fit value of \ww to any given model of quintessence based on  matching the evolution of the Hubble parameter $H(z)$ in a spatially flat universe. For the cases we tested, whose maps are shown in Fig. \ref{allLines}, we confirmed that the \ww parameterization was able to match the quintessence model's $H(z)$ with less than $0.7\%$ error at all $z$. We note that for each model considered in this work, the Swampland conjectures  \cite{Ooguri:2006in,Agrawal:2018own,Bedroya:2019snp} (thought to be required for a consistent theory of quantum gravity) are satisfied for a substantial range of parameters. The plateau and hilltop potentials, which  are sharply decreasing and can become negative, also have a natural place in cyclic bouncing cosmology \cite{Ijjas:2019pyf,Ijjas:2021zwv}.  

Though less precise than Bayesian model comparisons using Markov-Chain Monte Carlo simulations, this protocol provides a quick and useful way to visually compare different models of quintessence to each other and to observational data in a common two-dimensional parameter space. For example, if we take the constraints reported by DESI at face value, we see that plateau and hilltop models fare better than exponential models, with highly concave hilltop models being most compatible with the data. We reiterate, however, that this comparison test does not take into account any fine-tuning of the parameters or initial conditions of models that land within the observational constraint contours, and it is subject to the uncertainties discussed in Sec. \ref{uncertainty}.

This mapping protocol not only allows us to assess models of quintessence against observational data, but it also sheds new light on how to interpret the observational likelihood contours themselves. First, we have shown that the eccentricity and orientation of contours generated from measurements across a broad range of redshifts is a generic feature of the \ww parameterization. Second, we have found that for some models of thawing quintessence, the best-fit value of $w_0$ based on matching $H(z)$'s can differ significantly from the true value of $w_Q(0)$ predicted by the model. Finally, we have pointed out that the thawing quintessence models analyzed in this work, all of which obey the NEC (\emph{i.e.}, have $w_Q(z) \geq -1$), are mapped onto \ww combinations that satisfy $w_0 > -1$ and $w_0 + w_a < -1$. An observational preference for this sector, therefore, does not require the kinds of exotic field theories needed to enable a transition from NEC violation at large $z$ to NEC compliance at small $z$ (see, \emph{e.g.}, Ref.~\cite{Zhang:2005eg}).

This last finding has an important corollary: contrary to the suggestion in Ref. \cite{cortes_interpreting_2024}, we have shown that it is not just reasonable but crucially important for observational analyses to include combinations of \ww satisfying $w_0 + w_a < -1$ in their priors with high credence. Otherwise, these analyses would be inadvertently excluding families of simple, well-motivated models of thawing quintessence from consideration.

 \vspace{0.1in}
\noindent
{\it Acknowledgements.} 

This work is supported in part by the DOE grant number DEFG02-91ER40671 and by the Simons Foundation grant number 654561.  We thank  M. Ishak-Boushaki, S. Perlmutter and L. Page for useful comments on the manuscript.  PJS thanks Cumrun Vafa and the High Energy Physics group in the Department of Physics at Harvard University for graciously hosting him during his sabbatical leave.

\bibliographystyle{elsarticle-num.bst}
\bibliography{Assessing.bib}

\end{document}